\documentclass[aps,twocolumn,prb,superscriptaddress]{revtex4}

\usepackage{dcolumn}
\usepackage{amsmath}
\usepackage{graphics}

\begin{document}

\title {Spin-reorientation in YbFeO$_3$}

\author{Ya. B. Bazaliy}
  \affiliation{IBM Almaden Research Center, 650 Harry Rd., San Jose, CA 95120}
  \affiliation{O.Galkin Donetsk Physics \& Technology Institute,
  National Academy of Science of Ukraine, R.Luxemburg St. 72, Donetsk, 83114 Ukraine}
\author{L. T. Tsymbal}
  \email{tsymbal@sova.fti.ac.donetsk.ua}
  \affiliation{O.Galkin Donetsk Physics \& Technology Institute,
  National Academy of Science of Ukraine, R.Luxemburg St. 72, Donetsk, 83114 Ukraine}
  \affiliation{Ohio State University, Department of Physics,
  174 W. 18th Ave. Columbus, OH 43210}
\author{G. N. Kakazei}
  \affiliation{Ohio State University, Department of Physics,
  174 W. 18th Ave. Columbus, OH 43210}
  \affiliation{Institute of Magnetism, National Academy of Science of Ukraine, 36-B Vernadskii
Blvd., Kyiv, 03142  Ukraine}
\author{V. I. Kamenev}
  \affiliation{O.Galkin Donetsk Physics \& Technology Institute,
  National Academy of Science of Ukraine, R.Luxemburg St. 72, Donetsk, 83114 Ukraine}
\author{P. E. Wigen}
  \affiliation{Ohio State University, Department of Physics,
  174 W. 18th Ave. Columbus, OH 43210}

\date{\today}

\begin{abstract}
Precise measurements of YbFeO$_3$ magnetization in the spin-reoirentation temperature interval are performed. It is shown that ytterbium orthoferrite is well described by a recently developed modified mean field theory developed for ErFeO$_3$. This validates the conjecture about the essential influence of the rare earth ion's anisotropic paramagnetism on the magnetization behavior in the reorientation regions of all orthoferrites with $\Gamma_4 \to \Gamma_{24} \to \Gamma_2$ phase transitions.
\end{abstract}


\maketitle

\section{Introduction}

Rhombic rare-earth orthoferrites RFeO$_3$ with R being a rare-earth ion or an yttrium ion are magnetic insulators that provide a classic example of the second order orientation phase transitions. Orthoferrites have two magnetic subsystems: one of the rare-earth ions, and another of the iron ions. Magnetic properties of the subsystems and interaction between them depend on the external parameters, e.g. temperature, field, pressure, etc., and a series of phase transition is observed upon the parameter change.

In the temperature interval where phase transitions discussed in this work take place, the iron subsystem is ordered into a slightly canted antiferromagnetic structure exhibiting a weak ferromagnetic moment $\bf F$. The rare-earth system is paramagnetic. For all orthoferrites the antiferromagnetic structure below the Neel temperature $T_N$ ($T_N = 620 \div 740$K) corresponds to the $\Gamma_4$ ($G_x$, $F_z$) irreducible representation with magnetic vector $\bf F$ pointing along the $\bf c$ axis of the crystal and antiferromagnetic vector $\bf G$ pointing along the $\bf a$ axis. The coordinates are chosen so that ${\bf c} = \hat z$ and ${\bf a} = \hat x$. In orthoferrites with non-magnetic rare-earth ions (R = La, Lu, or Y) the $\Gamma_4$ ($G_x$,$F_z$) configuration persists to the lowest temperatures. For many other orthoferrites a reorientation transition with the sequence $\Gamma_4$ ($G_x$,$F_z$) $\to$ $\Gamma_{24}$ ($G_{xz}$,$F_{xz}$) $\to$ $\Gamma_2$ ($G_z$,$F_x$) is observed. Upon cooling vector $\bf F$ starts to rotate away from the $\bf c$ axis at temperature $T_1$. Its continuous rotation towards the $\bf a$ axis happens in the (${\bf a}, {\bf c}$) plane between temperatures $T_1$ and $T_2 < T_1$. Below $T_2$, the system stays in the $\Gamma_2$($G_z$,$F_x$) phase with $F || {\bf a}$.

Although the spin reorientation region $[T_2, T_1]$ has been studied for many orthoferrites by different experimental techniques, not enough is known about the specifics of the rotation. Relevant experimental results are often incomplete, lack accuracy, tend to contradict each other, and do not correspond to either conventional Landau theory~\cite{shane_varma,sirvardiere,belov_book} or its  suggested modifications. Recently~\cite{us_in_PRB,us_in_JAP} the temperature dependence of both $\bf a$ and $\bf c$ axis projections of the magnetic moment was measured with high accuracy for the single crystal samples of ErFeO$_3$. These measurements gave the temperature dependence of the absolute value of the magnetization $M(T)$ and its rotation angle $\theta(T)$ with respect to the $\bf c$ axis in the $[T_2,T_1]$ temperature interval at zero external magnetic field. The results were in very good agreement with the proposed modified mean field model,~\cite{us_in_PRB} that emphasized the anisotropy of the rare-earth ions paramagnetic susceptibility. It was conjectured that this model would be suitable for other magnetic materials with similar phase transitions. 

The present study is aimed at the detailed measurements of $M(T)$ and $\theta(T)$ behavior in single crystals of YbFeO$_3$, that exhibit the same $\Gamma_4 \to \Gamma_{24} \to \Gamma_2$
transition, with the purpose of checking this conjecture on another material. It is shown that the modified field theory of Refs.~\onlinecite{us_in_PRB,us_in_JAP} works well for YbFeO$_3$, even though in this orthoferrite the reorientation happens at an order of magnitude lower temperatures ($T \approx 8$K), than in ErFeO$_3$ ($T \approx 90$K), while the Neel temperature remains roughly the same $T_N \approx 630$K.

\begin{figure}[b]
\resizebox{.45\textwidth}{!}{\includegraphics{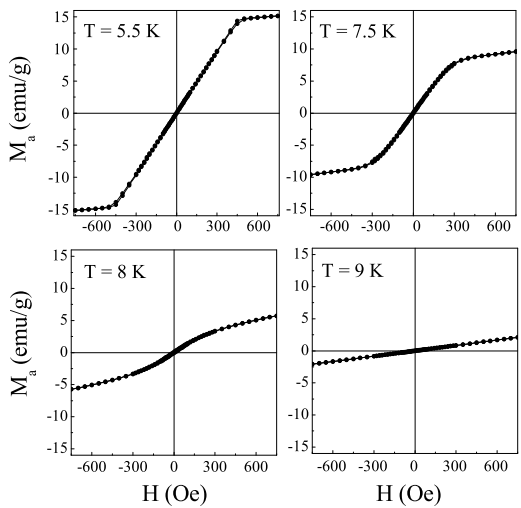}} 
\caption{Magnetization curves $M_a(H)$ obtained on the YbFeO$_3$, sample B, with $H||a$, at different temperatures.}
 \label{fig:1A}
\end{figure}

\begin{figure}[b]
\resizebox{.45\textwidth}{!}{\includegraphics{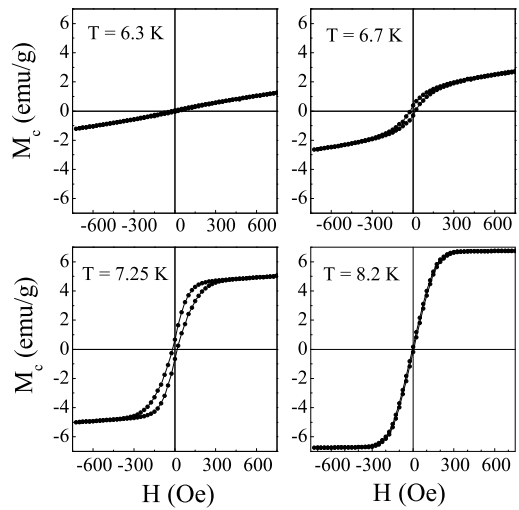}} 
\caption{Magnetization curves $M_c(H)$ obtained on the YbFeO$_3$, sample A, with $H||c$, at different temperatures.}
 \label{fig:1B}
\end{figure}

\section{Experimental results}

Measurements were performed on two single crystals of YbFeO$_3$. Cubic sample A, weighting 0.0485~g, was made of a single crystal grown by spontaneous crystallization in the melt-solution. Ellipsoid sample B, weighting 0.0715~g, was made of a single crystal grown by the no-crucible zone melting technique with radiation heating. The results for both samples are very similar. The temperature was varied in the 2 K to 10 K interval, and both $M_a$ and $M_c$ projections of the magnetic moment were
measured by a Quantum Design MPMS-5S SQUID magnetometer.

The ${\bf M}(T)$ dependence at zero external magnetic field was found through the analysis of magnetization curves analogous to those shown in Fig.~\ref{fig:1A},\ref{fig:1B}. To analyze the data we recall, that the $(H-T)$ phase diagrams in the vicinity of the $\Gamma_4 \to \Gamma_{24} \to \Gamma_2$ transition for ${\bf H} || {\bf c}$ and ${\bf H} || {\bf a}$ field directions are well known. According to them, as the magnetic field is swept through $H = 0$ inside the $[T_2, T_1]$ reorientation interval, a first order transition happens for both directions of the field and  manifests itself as a jump of magnetization component parallel to the applied field. First order transitions also happen above $T_1$ for ${\bf H} || {\bf c}$ and below $T_2$ for ${\bf H} || {\bf a}$ orientations, while no transitions are predicted below $T_2$ for $H || {\bf c}$ and above $T_1$ for $H || {\bf a}$. In a real experiment exact orientation of the field direction is obviously impossible. A three-dimensional diagram valid for the arbitrary field direction~\cite{bogdanov,us_in_PRB} shows that for a tilted field a first order transition happens at any temperature and a jump of at least one magnetic moment projection should be observed. In the case of single-domain switching that would produce a rectangular hysteresis loop.

Well-developed rectangular loops were indeed observed in experiments on ErFeO$_3$ outside of the reorientation interval.~\cite{us_in_PRB} Inside the $[T_2, T_1]$ interval ($T_1 \approx 88$K and $T_2 \approx 97$K for erbium orthoferrite) they transformed into the S-shaped magnetization curves. Such modification was attributed to the multi-domain state formation, possibly connected to the abrupt change in domain wall mobility.~\cite{rossol}

\begin{figure}[floatfix]
\vspace{-5mm}
 \resizebox{.4\textwidth}{!}{\includegraphics{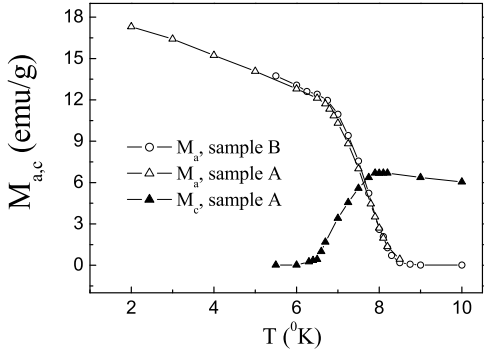}}
\vspace{-5mm}
\caption{Magnetization projections $M_{a,c}(T)$ obtained from the magnetization
curves: empty circles - $M_a(T)$ for sample B, empty triangles - $M_a(T)$ for
sample A, filled triangles - $M_c(T)$ for sample A.}
 \label{fig:2}
\end{figure}

In contrast with the case of ErFeO$_3$, magnetization curves in YbFeO$_3$ are S-shaped at all temperatures studied here for both ${\bf H} || {\bf a}$ and ${\bf H} || {\bf c}$ field orientations. The width of the magnetization curves for the magnetic field directed along the $\bf a$ axis is larger then for the field along the $\bf c$ axis. In general, the total width of the loops is considerably larger than in the case of ErFeO$_3$.~\cite{us_in_PRB} In accord with the phase diagrams discussed above, magnetization curves become straight lines passing
through the origin above $T_1$ for the ${\bf H} || {\bf a}$ orientation and below $T_2$ for the ${\bf H} || {\bf c}$ orientation. Their slope in these regions  corresponds to the paramagnetic contribution of the ytterbium ions (see Fig.~\ref{fig:1A},\ref{fig:1B}). Importantly, magnetization curves obtained for different samples are very similar. We extract the $\bf a$ and $\bf c$ projections of the bulk magnetization at zero external field by extrapolating their observed linear dependence at higher fields, $H \gtrsim 650$~Oe, by $M_{a,c}(H,T) = M_{a,c}(T) + \chi_{a,c}(T) H$ and extracting  the vertical intercept $M_{a,c}(T)$. The values of $M_a(T)$ and $M_c(T)$ obtained through this procedure are shown in Fig.~\ref{fig:2}. 

\begin{figure}[floatfix]
\vspace{-5mm}
 \resizebox{.4\textwidth}{!}{\includegraphics{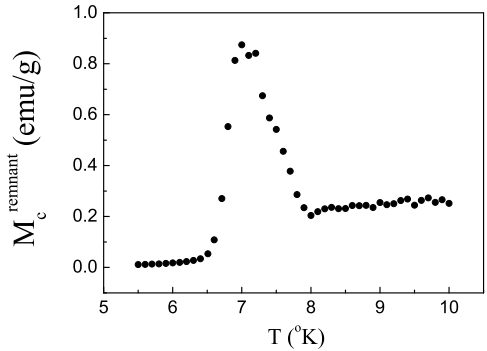}}
\vspace{-5mm}
\caption{Temperatue dependence of remnant magnetization along the $\bf c$-axis. Remnant magnetizations serves as a measure for the width of the hysteresis loops. Increase of hysteresis in reorientation region is clearly seen. The kinks of the curve mark the transition temperatures $T_{1,2}$}
\vspace{-5mm}
 \label{fig:hysteresis}
\end{figure}

It is interesting to note, that while being very small everywhere, the hysteresis of S-shaped loops visibly increases in the reorientation region of YbFeO$3$. This fact is illustrated by the following measurement. First, a saturating magnetic field $H = 650$Oe was applied either along the ${\bf a}$- or along the ${\bf c}$-axis. Then, the field was reduced to zero and the projection of the remnant magnetic moment ${\bf M}^{\rm remnant}$ on the same axis was measured. Two series of measurements, one for ${\bf M}^{\rm remnant}_c$ and another for ${\bf M}^{\rm remnant}_a$ were made. The results for the $H||c$ case are shown on Fig.~\ref{fig:hysteresis}. For S-shaped hysteresis loops the remnant magnetic moment grows with the width of the loop. The figure clearly demonstrates how the small hysteresis observed in the high-temperature symmetric phase $T \geq T_1$ grows inside the reorientation region and than drops to zero for $T \leq T_2$. Similar results were obtained for $H||a$. This property of hysteresis loops in YbFeO$_3$ turns out to be useful for the determinations of the critical temperatures $T_{1,2}$: they are clearly marked by the kinks of the curve on Fig.~\ref{fig:hysteresis} and give $T_1 = 8$K and $T_2 = 6.6$K.

The shape of the magnetization curves and the presence or absence of hysteretic
behavior depends on the quality of the samples, energy of the domain walls, etc.
The observed difference between erbium and ytterbium orthoferrites may result
from the order of magnitude difference in the temperature of the reorientation
transition. This question requires a separate study beyond the scope of the present paper. 

The absolute value of magnetization $M$ and rotation angle $\theta$ were extracted from the experimental data according to the expressions
\begin{eqnarray}
 \nonumber
 M &=& \sqrt{M_a^2 + M_c^2} \ , \quad
 \theta = \arctan \left( \frac{M_a}{M_c} \right)
\end{eqnarray}
and are shown in Fig.~\ref{fig:3} and Fig.~\ref{fig:4}. Experimental results
presented in Fig.~\ref{fig:2} and Fig.~\ref{fig:3} show that as the temperature is lowered from $T_N$ to $T_1$, the magnetization of the crystal gradually grows. This reflects the build up of the iron moment near $T_N$ and subsequent development of the ytterbium moment along iron moment~\cite{white_JAP} at lower temperatures. In the narrow reorientation region $[T_2, T_1]$ the magnetization rapidly grows almost two-fold. Below $T_2$ the magnetization continues to grow. This supports the result of Ref.~\onlinecite{bozorth}, suggesting that the ytterbium moment remains parallel to the iron moment, and does not switch to the antiparallel direction as stated in Ref.~\onlinecite{white_JAP}.

\section{Theoretical analysis}

\begin{figure}[floatfix]
 \resizebox{.45\textwidth}{!}{\includegraphics{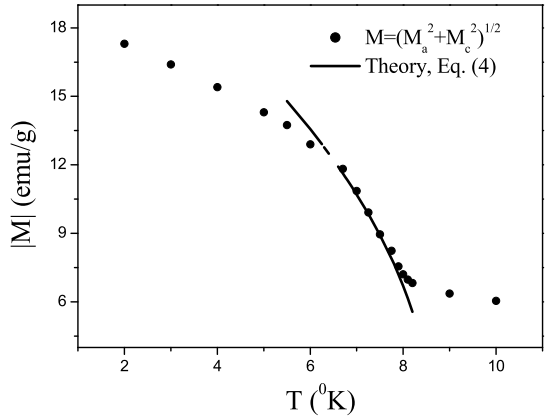}}
\caption{Absolute value of the magnetization $M(T)$ calculated from experimental
data. Solid curve - theory Eq.~(\ref{eq:4})}
\vspace{-5mm}
 \label{fig:3}
\end{figure}

\begin{figure}[floatfix]
 \resizebox{.45\textwidth}{!}{\includegraphics{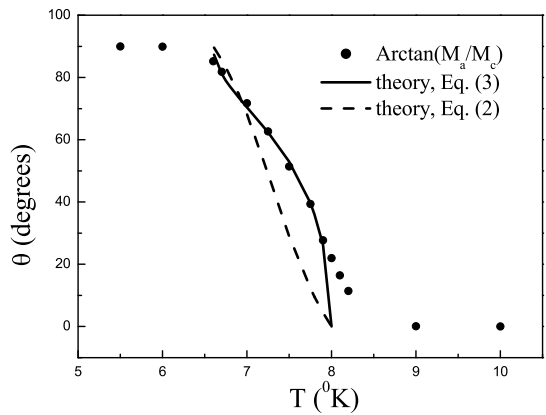}}
\caption{Magnetization rotation angle $\theta(T)$ calculated  from experimental
data in the reorientation region $[T_2, T_1]$  at zero external magnetic field.
Solid curve - theory Eq.~(\ref{eq:3}), dash curve - conventional theory
Eq.~(\ref{eq:2})}
 \label{fig:4}
\end{figure}

Our experimental results can be explained by the modified mean field theory suggested in Refs.~\onlinecite{us_in_PRB,us_in_JAP}. As the conventional Landau theory,~\cite{shane_varma,sirvardiere,belov_book} the modified theory assumes that the magnetization of iron subsystem is saturated at $T \lesssim T_{1,2} \ll T_N$. The free energy of the iron subsystem is taken in the form
\begin{eqnarray}
 \label{eq:1}
 F(\theta,T) &=& F_0(T) + \frac{K_u(T)}{2} \cos(2\theta) + K_b\cos(4\theta) \ .
\end{eqnarray}
With minimal assumptions about the temperature dependence of phenomenological
constants inside the reorientation region, namely constant $K_b$, and $K_u(T)$
linearly varying with temperature and going through zero inside the
reorientation interval, the minimization of the conventional energy functional
(\ref{eq:1}) gives ~\cite{shane_varma,sirvardiere,belov_book}
\begin{eqnarray}
 \label{eq:2}
 \tan \theta &=& \sqrt{\frac{1+\xi}{1-\xi}} \ ,
 \quad
 \xi(T) = \frac{(T_1 + T_2)/2 - T}{(T_1 - T_2)/2}
\end{eqnarray}
Figs.~\ref{fig:3}, \ref{fig:4} show that experimental results neither support
the constancy of $M(T)$, nor give a $\theta(T)$ dependence consistent with
Eq.~(\ref{eq:2})

According to the modified mean field model, paramagnetic susceptibility of ytterbium subsystem should also be taken into account to adequately describe the magnetic behavior of the orthoferrite. It is assumed, that in the molecular field of iron the rare-earth ion acquires a magnetic moment ${\bf m} =
\hat\chi^{\rm Yb} {\bf F}$, while the absolute value of the iron moment $\bf F$ remains constant.~\cite{sirvardiere,white_JAP,treves,yamaguchi} Experimentally measured magnetization is the sum of the iron and rare-earth contributions ${\bf M} = {\bf F} + {\bf m}$. The magnetic susceptibility $\hat\chi^{\rm Yb}$ of the  rare-earth ions is assumed to be anisotropic. This assumption naturally explains the large change of $M$ inside a narrow temperature interval, since rotation of $\bf F$ leads to the change of $\bf m$ and thus changes $M$ as well.\cite{us_in_PRB} The anisotropy of the rare-earth susceptibility has been reported in the literature.~\cite{sirvardiere,white_JAP,treves,yamaguchi} The key point of Ref.~\onlinecite{us_in_PRB} was the proper account of such anisotropy in the calculation of the rotation angle and absolute value of the magnetization, with the result:
\begin{eqnarray}
 \label{eq:3}
 \tan \theta &=& r \sqrt{\frac{1+\xi}{1-\xi}} \ , \quad
 r = \frac{M_a(T_2)}{M_c(T_1)}
 \\
 \label{eq:4}
 M &=& M_c(T_1) \sqrt{\frac{r^2(1+\xi) + (1-\xi)}{2}} \ .
\end{eqnarray}
Since $M_a(T_2)$  and $M_c(T_1)$ are measurable magnetizations of the sample at
temperatures $T_2$ and $T_1$ respectively, the value of $r$ is known and
expressions (\ref{eq:3}) and (\ref{eq:4}) have no fitting parameters. 

According to our measurements, the cubic sample A, for which most of the measurements were done, had $T_1 = 8.0$ K and $T_2 = 6.6$ K. Using the values of $M_a(T_2)$ and $M_c(T_1)$ at these temperatures we find $r = 1.78$. Theoretical curves given by Eqs.~(\ref{eq:3}) and (\ref{eq:4}) are shown in Figs.~\ref{fig:3} and \ref{fig:4} by solid lines. A convincing correspondence between the theory and experiment is evident. 

Note, that the analysis of experimental data in terms of the model introduced in Refs.~\onlinecite{us_in_PRB,us_in_JAP} is only valid inside the reorientation region. However, it is important that inside the region of its validity such analysis is independent of the driving mechanism of the transition, be it the interactions in the iron subsystem, the R-Fe interactions, the behavior of the rare-earth magnetic succeptibility, or any other process. The approach of Refs.~\onlinecite{us_in_PRB,us_in_JAP} only requires the effective anisotropy constant $K_u(T)$ to be a linear function of temperature. Since precisely that behavior of $K_u(T)$ was measured in Refs.~\onlinecite{abe,belov}, and Ref.~\onlinecite{schaffer} shows that such behavior follows from the microscopic model of Ref.~\onlinecite{levinson}, the modified mean field theory \cite{us_in_PRB,us_in_JAP} can be applicable for a wide variety of orthoferrites.

\bigskip

\section{Conclusion}

In this paper we report direct measurements of the magnetization aboslute value $|{\bf M}|(T)$ and rotation angle $\theta(T)$ during the $\Gamma_4 \to \Gamma_{24} \to \Gamma_2$ spin-reorientation transition in YbFeO$_3$ single crystals. The results favor the importance of strongly anisotropic rare-earth contribution to the magnetization of the material. They give a convincing argument if favor of the spin reorientation model suggested in Ref.~\onlinecite{us_in_PRB} and its applicability to $\Gamma_4$ ($G_x$,$F_z$) $\to$ $\Gamma_{24}$ ($G_{xz}$,$F_{xz}$) $\to$ $\Gamma_2$ ($G_z$,$F_x$) orientation transitions in different materials.

\section{Acknowledgements}

The work at O.Galkin Physics \& Technology Institute was partially supported by
the State Fund for Fundamental Research of Ukraine, project F7/203-2004. Ya.B.
was supported by DARPA/ARO, Contract No.~DAAD19-01-C-006.

\end{document}